# New concept for the development of Bi-2212 wires for high field applications

A Malagoli[1], A Leveratto[1], V Braccini[1], D Contarino[1,2], and C Ferdeghini[1]

[1] CNR – SPIN Genova, Corso Perrone 24, 16152 Genova, Italy

[2] DIFI, University of Genova, Via Dodecaneso 33, 16145 Genova, Italy

E-mail: andrea.malagoli@spin.cnr.it

**Abstract**

The first step towards high critical currents in Bi-2212 wires was the comprehension that the supercurrent is blocked over long lengths by filament-diameter bubbles grown during the melt stage, which cause expansion of the wire diameter and dedensification of the superconducting filaments. Whereas the previous successful approach to reduce the problem of voids related to bubbles was based on the application of a high overpressure during the heat treatment, we fabricated Bi-2212 wires by applying a new concept of suitably alternating groove-rolling and drawing techniques with the aim of densifying the phase already during the working procedure prior to the heat treatment.

We here for the first time were able to reach in wires reacted with closed ends – i.e. with gas trapped in the wire as it happens in long-length wires - the very same values of critical current shown in short wires reacted with open ends. This is the irrefutable evidence that, only by acting on the deformation technique, we were able to raise the critical current by properly densifying the superconducting powder inside the filaments already before the melt stage. Whole-conductor current densities in our long length simulation wires already reach 400 A/mm$^2$ at 4.2 K and 5 T, which can be still easily increased through architecture optimization. The actual breakthrough is that the densification is optimized without further complex treatments through a technique which can be straightforwardly applied to long-lengths wires.

# 1. Introduction

Bi-2212 recently hit the headlines in the applied superconductivity community after it was demonstrated that it possesses the highest engineering current density $J_E$ of any available conductor above 17 T at liquid Helium temperature when treated overpressure[1], especially this being achieved in the much preferable isotropic, round-wire architecture and multifilamentary form. These two simultaneous characteristics make Bi-2212 particularly appealing for high-field magnets for applications such as MRI or NMR magnets, particle accelerators or fusion devices.

For long time, high critical current density ($J_c$) in Bi-2212 could be obtained only in short wires, until it has been understood that the main obstacle to current flow through the Bi-2212 filament is given by the presence of bubbles formed during the melt step of heat treatment [2, 3]: the pressure caused by them provokes a Bi-2212 movement and an expansion of the wire during the melt stage[4] that makes $J_c$ very inhomogeneous over lengths longer than 10-15 cm [3]. Different approaches have been developed in order to increase the density of the filaments in the commercial Oxford Superconducting Technology (OST) wires and therefore to diminish the bubble density and/or size. They all act on the final as-drawn conductor: $J_c$ was first doubled by using a 2 GPa cold isostatic pressure (CIPping) [5] or through swaging [6] after drawing, while very high over pressure (OP) applied during the heat treatment was able to strongly increase the connectivity of the Bi-2212 phase and, for OP of 100 bar, to raise $J_E$ by up to 8 times reaching almost 1000 A/mm$^2$ at 4.2 K and 5 T in wires denser than 95% [1]. Such value was measured in short samples (~ 8 cm) reacted with sealed ends, which are considered a good simulation of long length wires, where the gas generated internally remains trapped in the wire [3]. A test coil was also prepared from 30 m of Bi-2212 wire and reacted at 10 bar overpressure: it generated 2.6 T in a background field of 31.2 T and showed a $J_E$ of 360 A/mm$^2$ at 5 T, the highest value measured in an actual long length wire [1]. Up to now, the OP process was believed to be the only chance for Bi-2212 to be applied.

With the same aim of raising the critical current in long length wires enhancing the Bi-2212 density in the conductor and overcoming the formation of bubbles, we adopted a different approach: instead of developing treatments on the commercially available as-deformed wire, we started from the fabrication of the wire itself acting on the deformation process with the aim to fabricate a denser wire – keeping in mind as a key point its straightforward scalability over industrial lengths. In particular, we applied the groove-rolling instead of the 'standard' drawing procedure: the basic concept lies in the fact that in the drawing, the strain is shared in a compressive transversal component and in a remarkable stretching longitudinal component, while when deforming through groove-rolling the strain is almost completely in the compressive transversal direction. Such a technique has greater powder compaction power, once the wire is optimized it is straightforwardly adaptable to long length wires to make coils, and allows the fabrication of samples with not only round, but also square or rectangular shape depending on the application requirements. This last characteristic can present some advantages: in a solenoid type winding as NMR magnets, for example, the rectangular shape allows a better compaction of the turns drastically reducing the voids space between them and avoids unwanted conductor twisting during the winding process; nevertheless the corners are smooth and this is a very desirable condition to avoid damages in the insulation, and the geometric ratio stays around 1.5, therefore maintaining the peculiar Bi-2212 isotropy.

In our previous publications [7, 8], we demonstrated the ability of groove-rolling to increase the density in Bi-2212 wires, leading to a threefold increase in $J_c$ with respect to drawn wires in square and rectangular wires with various architectures. $J_c$ up to 2000 A/mm$^2$ were measured in open ends wires, but the remarkable result was the reduction of only 28% between open- and closed-ends wires which is much lower than the 70% reduction observed in the commercial OST wires, clear indication of a higher density reached through groove-rolling [8].

In this paper we report about a new concept of combining in a proper way the peculiar characteristics of both deformation processes, i.e. the capability to compact the Bi-2212 particles proper of groove-rolling with that of making slide them inside the tube, peculiar of the drawing, with the final aim to get the best powder packing and increase the density of the superconducting phase already before the melt stage. In our process - which we will indicate as GDG process in the following - the two techniques are alternated in a suitable way: after the grove-rolling stages compact the powders as the best of its possibility in the transversal direction, the drawing steps make them flow so as to fill in the voids in the longitudinal direction, and so on. We focused in particular on the behavior of samples heat treated at 1 bar with ends either open or closed trying to rise the performances of the sample with closed ends up to those required by the applications. The optimal alternation of groove-rolling and drawing steps gave as a result wires which for the first time do not show any degradation when the ends are sealed, reaching $J_E$ of 400 A/mm$^2$ at 4.2 K and 5 T without application of any expensive and not easily scalable OP process.

**2.    Experimental details**

Bi-2212 multifilamentary wires were prepared through the Powder-In-Tube (P.I.T.) technique. Ag tubes – with outer (OD) and inner (ID) diameters of 15 and 11 mm respectively - were first filled with Nexans granulate powder - the same used by OST for the production of their commercial wires - whose overall composition was $Bi_{2.16}Sr_{1.93}Ca_{0.89}Cu_{2.02}O_x$. After drawing, the obtained monofilamentary wire was hexagonal shaped and cut in 85 pieces, restacked in a second 12.5/11 (OD/ID) mm Ag tube and then cold worked with a proper alternation of drawing and groove-rolling steps. The obtained wire was again hexagonal shaped, cut in 7 pieces and restacked in a 9.5/8 mm (OD/ID) Ag/Mg alloy tube. The restacked tube underwent again a process with alternation of drawing and groove-rolling. Finally, a rectangular wire 1.15 x 0.7 mm$^2$ (R) and a square wire 0.7 x 0.7 mm$^2$ (S) were obtained, both having a superconducting fill factor of about 18%. In Figure 1 we report the cross sections of both wires.

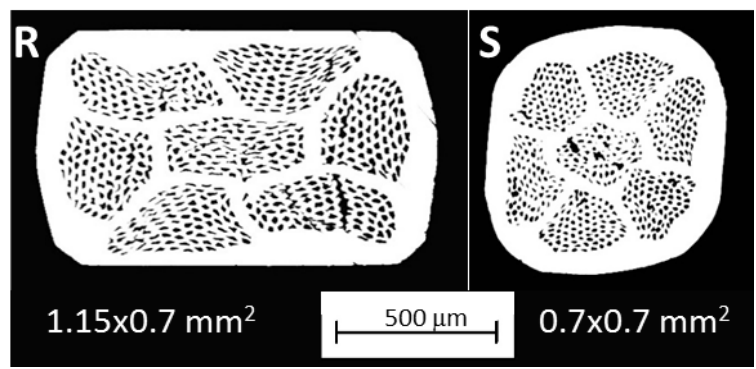

**Figure1.** Images of the transverse cross sections of the Rectangular (R - left panel) and Square (S – right panel) multifilamentary wires, both having a fill factor of about 18%.

12 cm long samples cut from each wire were heat treated in 1 bar flowing $O_2$ in a tubular furnace with a homogeneity zone (±0.5 °C) of 30 cm using the standard heat treatment (HT) schedule [9]. Both samples were treated with closed (R_CE, S_CE) and open (R_OE, S_OE) ends: CE ends samples were sealed by dipping them into liquid silver before the heat treatment, and after the heat treatment no leakages, cracks, diameter expansions or bubbles were observed neither along the wire nor at their edges. Longitudinal cross sections were prepared through removing the silver matrix by deep chemical etching using a solution based on ammonium hydroxide (50 vol% aqueous solution) and hydrogen peroxide (30 wt% aqueous solution), mixed in equal amounts, partly diluted by distilled water, in order to expose the Bi-2212 filaments. Microstructures were analyzed by scanning electron microscopy (SEM).

Transport critical currents ($I_c$) were measured on 12 cm long samples by means of a four-probe system in a 7 T split-coil magnet at 4.2 K with the field applied perpendicular to the wire axis. The voltage taps were located about 1 cm apart in the center of the 12 cm long samples; a criterion of 1 μV/cm was used. The critical current density $J_c$ was calculated taking into account the superconducting filament area measured by image analysis on the unreacted wire, as it is used in the literature [10], while the engineering current density $J_E$ was obtained simply dividing by the overall wires cross-section.

## 3. Results and discussion

In Figure 2, $J_c$ (left axis) and $J_E$ (right axis) are shown for the two samples, either with closed and open ends. The most striking result is that for the first time, and this property was confirmed on several couples of wires prepared in the same way, we were able to measure identical critical current values in wires heat treated with open- and closed-ends. We had already shown that the groove-rolling process was able to densify the superconducting powders and therefore diminish the bubble density: the fact that we now register no reduction at all in the simulation of long length samples is the confirmation that it is possible, only acting on the deformation process in a proper way, to reach an optimal densification of powders inside the filaments and therefore completely overcome the problems generated by the bubbles inside the wires. In practice, the pressure applied during the deformation is enough to prevent the bubble agglomeration as well as the deleterious creep dedensification driven by the internal gas pressure during the heat treatment.

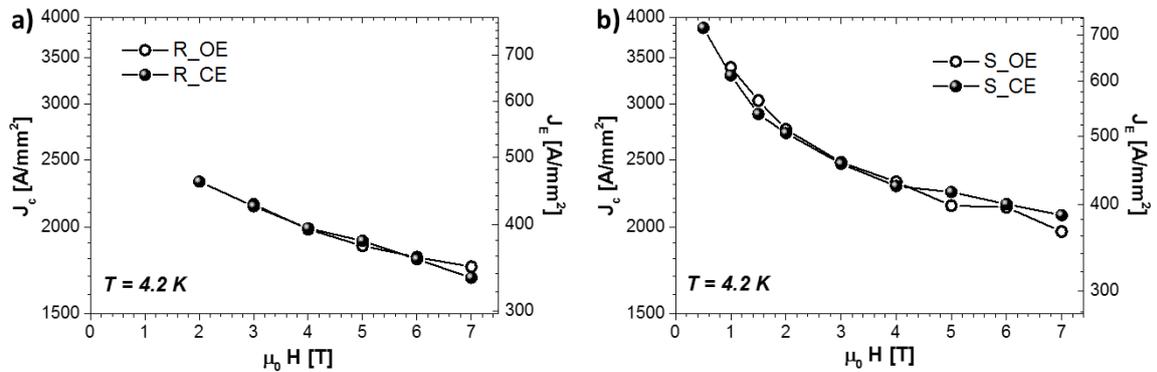

**Figure 2.** $J_c$ (left axis) and $J_E$ (right axis) for the a) rectangular R and b) square S samples with both closed (CE) and open (OE) ends

In Figure 3 we report the longitudinal cross sections of the square wire at the end of the deformation process and after the full heat treatment, in both cases compared with the longitudinal cross section of a round wire prepared only through drawing, with a diameter of 0.8 mm and the same internal architecture [7]. Filaments size is about the same for the two wires. From the images of the green wires we actually observed what we hypothesized at the beginning and described earlier: in the case of drawing alone, the particles are clearly distributed on longitudinal lines, while in the wire prepared with the alternation of groove rolling and drawing the particles look much more compacted already before the heat treatment, as if after flowing in the longitudinal direction the voids (visible in the left figure) were filled thanks to a compression in the transversal direction due to the groove-rolling. Also the images after the melt are quite meaningful: the drawn sample, exactly as it happens in the commercial wires [1], presents many untextured regions, the filaments can be barely distinguished and many voids are visible, while the wire prepared with our new method shows a higher texturing degree with filaments certainly denser. In particular, grains grown in direction perpendicular to the wire, which obstruct the passage of the current in a particular filament and are quite easily found in drawn wires, are almost absent in GDG samples.

Above all that, $J_E$ values of 400 A/mm$^2$ at 5 T in wires with closed ends which did not undergo any further treatment are already truly remarkable as absolute values, corresponding to the $J_E$ measured on a OST wire treated with an over pressure of 10 bar and being comparable with the $J_E$ of 360 A/mm$^2$ of the 30 m long test coil reacted in a 10 bar overpressure [1]. They open up very interesting perspectives for the fabrication of superconducting windings realized with long-length wires with the desired shape, without problems due to Bi-2212 leakage after melt processing and without the need of using complex and expensive treatments.

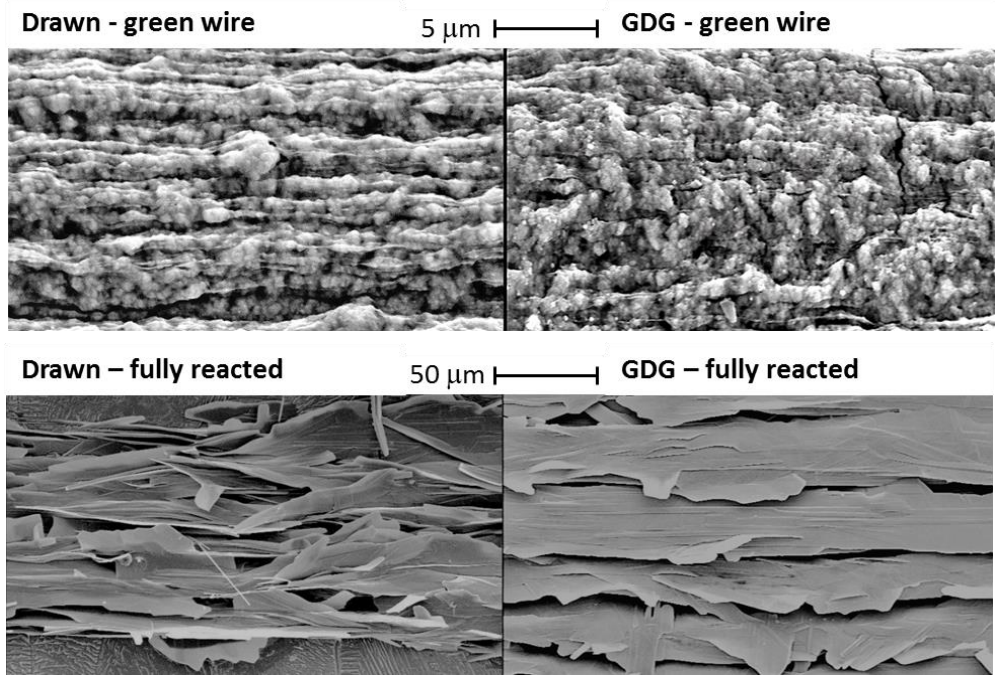

**Figure 3.** SEM images of the longitudinal cross sections of the wire prepared through drawing (left side) and through our GDG process described in this paper (right side), either before (upper panels) and after (lower panels) the full heat treatment.

Regarding the absolute values of $J_c$, which is a more intrinsic property, the rectangular wire shows about 1900 A/cm$^2$ at 5 T which reaches 2200 A/cm$^2$ in the square sample: such values are four times higher than the $J_c$ values for the commercial round Bi-2212 wires heat treated at the same room pressure [1], and this improvement is entirely due to the mechanical process. As it is possible to observe in Figure 1, the filaments are quite irregular in shape, especially in the rectangular sample where they appear particularly straightened. We tried to measure the average dimensions of the filaments and we estimated their size around 20 μm for the rectangular wire and 14 μm for the square sample. This explains the better performances of the square sample in terms of $J_c$, as previous studies have shown how 15 μm is the value for the optimum $J_c$ performance [11, 12].

As mentioned above, both multifilamentary wires have a superconducting cross section of only about 18%, well below the 28% of the commercial OST wires. We therefore still have a strong margin for improvement: we need to optimize the wire in terms of geometry of the restacks, thickness of the external and internal Ag tubes, filament average dimensions, and in general filling factor. By increasing the fill factor from 18 to 28% we expect a straightforward improvement of the overall conductor density, which could reach values above 600 A/mm$^2$, value which is estimated as a suitable value for high field applications [13], even more in wires prepared with a process which can be applied to long length wires and eventually industrialized in a quite immediate way, making it particularly appealing being faster and cheaper than any other.

## 4. Conclusions

In summary, we report about a new concept for the fabrication of Bi-2212 wires which can reach critical current values useful for high field applications also in long lengths. We developed a new process where drawing and groove-rolling procedures are conveniently alternated in order to get the best powder packing already before the melt stage. In particular, we were able to fabricate wires which for the first time do not show any degradation when the ends of the wires are sealed – which is considered a reliable approximation of long-length wires - reaching engineering current densities of 400 A/mm$^2$ at 4.2 K and 5 T without the application of any over pressure. Such values can be still easily increased through architecture optimization.

The actual innovation is that the densification is optimized already before the melt stage without further treatments and through a technique which can be straightaway exploited on long-lengths wires.

## Acknowledgements

We thank C. Bernini for the SEM images and I. Pallecchi for fruitful scientific discussion.